\documentclass[aps,prl,twocolumn,superscriptaddress,nofootinbib]{revtex4-2}

\usepackage{amsmath,amssymb,amsfonts}
\usepackage{bm}
\usepackage{graphicx}
\usepackage{booktabs}
\usepackage{hyperref}
\usepackage{xcolor}
\usepackage{tikz}
\usepackage{float}
\begin{document}

\title{Scaling law of variational quantum algorithms}

\author{Ningfeng Wang}
\email{wnf22@mails.tsinghua.edu.cn}
\affiliation{Yau Mathematical Sciences Center, Tsinghua University, Beijing 100084, China}
\affiliation{Yanqi Lake Beijing Institute of Mathematical Sciences and Applications, Beijing 101408, China}
\author{Zhengwei Liu}
\email{liuzhengwei@tsinghua.edu.cn}
\affiliation{Yau Mathematical Sciences Center, Tsinghua University, Beijing 100084, China}
\affiliation{Yanqi Lake Beijing Institute of Mathematical Sciences and Applications, Beijing 101408, China}

\date{\today}
\begin{abstract}
We study scaling law of variational quantum algorithms by large-scale numerical experiments on combinatorial optimizations. On experiments up to $n=5000$ qubits with $\text{O(}\log(n))$ circuit depth with a classically simulable ansatz, the converged energies consistently differ by no more than 0.5\% from the optimal or reference energies reported in the benchmarks.
\end{abstract}
\maketitle

Rapid progress in quantum technologies is driving near-term quantum devices beyond the intermediate scale \cite{Preskill2018NISQ} of a hundred toward large-scale implementation of thousands of physical qubits. Although the scale remains insufficient for fault tolerant quantum computations, it is of great interest to develop near-term quantum advantages with useful applications.

A proper candidate approach towards near-term practical quantum advantage is variational quantum algorithm \cite{Cerezo2021}. For instance, quantum approximate optimization algorithms (QAOA) and the variational quantum eigensolvers (VQE), encode a classical objective into a implementable cost Hamiltonian and optimize the expectation value over a parameterized quantum state. However, the effectiveness with scaling law had been unclear due to technical obstacles. Physically motivated and structured ansätze, such as unitary coupled-cluster singles and doubles (UCCSD) \cite{McClean2016}, may require deep circuit depth and are not implementable on near-term quantum hardwares. By contrast, hardware-efficient ansätze \cite{Kandala2017}, when sufficiently deep and randomly initialized, may approach unitary 2-designs, resulting in concentration of measure and exponentially vanishing gradient variances—a phenomenon known as a barren plateau \cite{McClean2018,liu2022presence}, as a consequence of concentration of measure over an exponentially large reachable space associated with the circuit’s dynamical Lie algebra \cite{Larocca2022, Ragone2024, allcock2026dynamical}, which requires special techniques to repair \cite{chen2025taming}. In addition, gradient estimation on quantum hardware requires many circuit evaluations and suffers from sampling noise and device noise. Due to currently limited scale of quantum computers and classical simulations, the scaling law of variational quantum algorithms had been unrevealed.

In this article, we reveal the scaling law of variational quantum algorithms, on numerical experiments of up to 5000 qubits with logarithmic circuit depth on a classically simulated 1D ansatz, and with near logarithmic gradient counts, which gives effective results by convergence exceeding 99.5\% of optimal or reference energies of various combinatorial optimization and NP complete problems. Combined with post-processing of these high-quality solutions, the energies are further improved. On near-threshold 3-SAT problems, there are high chances of finding solutions as global optima.  \begin{figure}[H]
    \centering
    \includegraphics[width=1\linewidth]{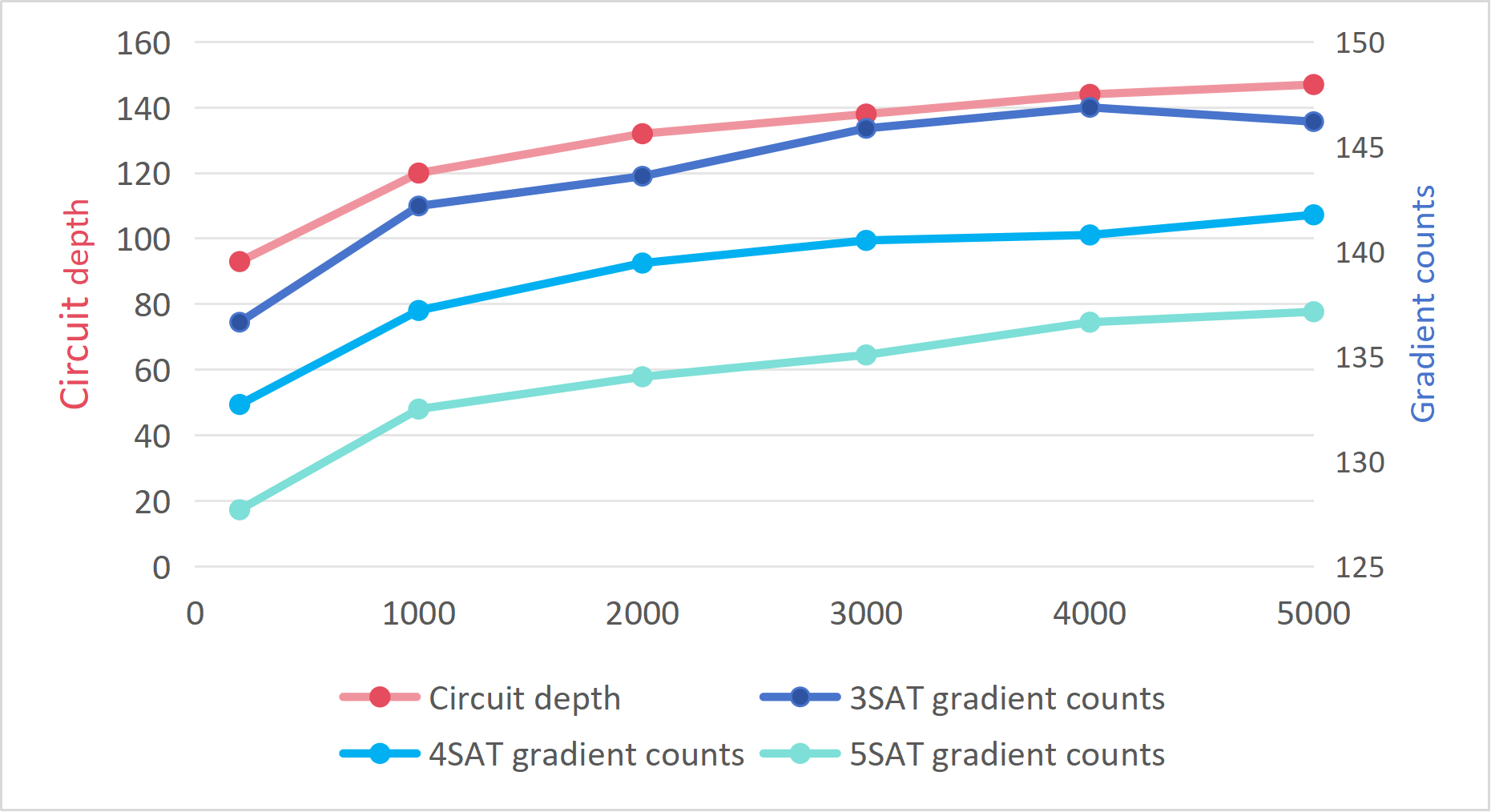}
    \caption{Scaling law: k-SATs. The chart is based on Table II III IV.}
    \label{fig:ksat-scaling}
\end{figure}

To study the scaling law on various practical problems, we adopt a general architecture for variational quantum algorithms, formulated as Boolean combinatorial optimization problems, which is central to theoretical computer science, operations research, artificial intelligence and broad fields of finance, transportation and engineering. Canonical examples include maximum independent set, graph coloring, quadratic unconstrained binary optimization (QUBO) and maximum satisfiability (max-SAT). Due to their discrete and typically nonconvex structure, these NP-hard problems provide a natural testing ground for heuristic algorithms for both classical and quantum methods. In particular, many such problems can be formulated as QUBO or Ising
models and are therefore directly amenable to implementation on quantum
annealing platforms \cite{Johnson2011DWave,Boixo2014QuantumAnnealing} and coherence Ising machines \cite{wei2026versatile}. 

To implement on variational quantum algorithms, boolean assignments are encoded in the computational basis, while constraints, penalties, rewards, and objectives are encoded as diagonal cost Hamiltonians. A parameterized quantum circuit thereby defines a trainable probability distribution over Boolean strings, and energy minimization drives sampling toward high-quality solutions. The framework is ansätze-independent and provides a unified route for applying variational quantum algorithms to SAT, Max-SAT, CSP, QUBO, graph optimization, and other Boolean optimization problems. Recent work ~\cite{lu2026evidence} has reported numerical evidence of a scaling advantage over classical baselines for enhanced quantum solvers on instances of the NP-complete one-in-three SAT problem with up to 70 variables. 

To investigate the scaling law of the general framework on general quantum solvers, we instantiate the architecture with a problem-independent classically simulable matchgate ansätze \cite{Valiant2002, Cai2009Matchgate}. Owing to its equivalence to fermionic Gaussian \cite{Bravyi2005} evolution, this realization admits polynomial-time classical simulation through Majorana covariance matrices, enabling deterministic evaluation of losses and gradients without exponential state-vector storage. This ansatz allows us to test the scaling law of variational quantum algorithms at a 5000-qubit scale adopting only logarithmic circuit depth, far beyond capacity of conventional state-vector simulation. On planted Boolean optimization instances, including QUBO and planted critical k-SAT instances for $k=3,4,5$ with up to 5000 variables, the method consistently converges to near-optimal energies exceeding 99.5\% of the optimum. These results show that the variational Boolean-optimization architecture is trainable at large scales and that the matchgate realization provides an efficient testbed for assessing quantum machine-learning approaches before deployment on more expressive quantum hardwares. 
\paragraph{Methods.}
We consider Boolean optimization problems of the form
\[
\min_{x\in\{0,1\}^n} f(x),
\]
where $x=(x_1,...,x_n),x_i\in \{0,1\},$ $f:\{0,1\}^n\rightarrow \mathbb{Z}_2 \text{ or }\mathbb{R}$ is a Boolean or pseudo-Boolean function. 

Each Boolean variable is encoded into a qubit, and all instances form a computational basis of n-qubit Hilbert space $\mathcal{H}=\text{span}\{0,1\}^n$. The classical cost function is mapped to a diagonal Hamiltonian
\[
H_f=\sum_{x\in\{0,1\}^n} f(x) |x\rangle\langle x|.\]

Therefore, we perform optimizations of the following problem on a quantum state:
\[
\min_{|\psi\rangle \in B^1(\mathcal{H})} \langle \psi|H_f|\psi\rangle,
\]
where $B^1(\mathcal{H})$ denotes the n-qubit Bloch ball, the unit ball of $\mathcal{H}$. 

For one way of implementation on near-term quantum computers, the Hamiltonian can be decomposed into polynomial scale Pauli words. For example, we set $f$ to be a polynomial scale summation of $m=\text{poly}(n)$ sparse terms: 
$$f(x)=\sum_{j=1}^m w_j f_j(x), $$
where each term \(f_j(x)\) is a Boolean or pseudo-Boolean function, representing a constraint violation penalty or a reward term that is $K$-sparse, i.e. $f_j(x_{i_1},x_{i_2},...,x_{i_k}), k\textless K, K\ll n$ involving less than $K$ variables. This formulation covers Boolean polynomial equations, inequalities, modular constraints, logical clauses, graph conflict penalties, vertex-selection rewards, and other local pseudo-Boolean terms. Therefore, the Hamiltonian can be decomposed into sparse terms:
$$H_f=\sum_{j=1}^m H_{f_j}, H_{f_j}=\sum_{x\in\{0,1\}^n} f_j(x) |x\rangle\langle x|.$$
Each $H_{f_i}$ can be decomposed into at most $2^K$ Pauli-$Z$ words since $f_j$ involves at most $K$ variables.

A parameterized quantum state
\[
|\psi(\theta)\rangle=U(\theta)|00...0\rangle
\]
defines a probability distribution
\[
p_\theta(x)=|\langle x|\psi(\theta)\rangle|^2.
\]
The training loss is the expected cost
\[
L(\theta)=\langle\psi(\theta)|H_f|\psi(\theta)\rangle
=\sum_x p_\theta(x)f(x).
\]
Thus quantum superpositions encodes discrete combinatorial decisions into continuous optimizations, and training reshapes the sampling distribution toward low-cost Boolean strings.

To construct the Hamiltonian $H_f$, for a local Boolean function \(f_j(x_{i_1},\ldots,x_{i_k})\), the corresponding diagonal observable can be obtained by replacing
\[x_i\mapsto \hat{x}_i=\frac{I-Z_i}{2}.\]
Equivalently, computational-basis projectors are written as
\[P_i^{(0)}=\frac{I+Z_i}{2},\qquad
P_i^{(1)}=\frac{I-Z_i}{2}.\]
Thus a local truth-table term can be represented as a linear combination of \(Z\)-strings.

As a simple example, consider three Boolean variables
\(x_1,x_2,x_3\in\{0,1\}\) with two constraints
\[
f_1(x)=x_1(1-x_2)=0,\qquad
f_2(x)=x_2\oplus x_3=1.
\]
For each constraint we assign energy \(-1\) if it is satisfied and
\(+1\) if it is violated.  Equivalently, if \(V_j(x)\in\{0,1\}\)
denotes the violation indicator, we set
\[
f_j(x)=2V_j(x)-1.
\]
For the first constraint,
\[
V_1(x)=x_1(1-x_2),
\]
and therefore
\[
H_{f_1}
=
2\frac{I-Z_1}{2}\frac{I+Z_2}{2}-I
=
\frac{1}{2}\left(I+Z_2-Z_1-Z_1Z_2\right)-I.
\]
For the second constraint, violation means \(x_2\oplus x_3=0\), hence
\[
V_2(x)=1-(x_2\oplus x_3)=\frac{1+Z_2Z_3}{2}
\]
at the operator level, and
\[
H_{f_2}
=
2V_2-I
=
Z_2Z_3.
\]
The total diagonal Hamiltonian is
\[
H_f= H_{f_1}+ H_{f_2}.
\]

\emph{Matchgate ansätze.}
We choose \(U(\theta)\) from the matchgate family, or fermionic Gaussian unitaries. For \(n\) qubits, define Majorana operators
\[
c_{2j-1}=Z_1Z_2\cdots Z_{j-1}X_j,
\]
\[
c_{2j}=Z_1Z_2\cdots Z_{j-1}Y_j.
\]
A fermionic Gaussian unitary satisfies
\[
U^\dagger(\theta)c_pU(\theta)=\sum_{q=1}^{2n}R_{pq}(\theta)c_q,
\]
where \(R(\theta)\) is a real orthogonal matrix. This condition defines the Majorana linear transformation property.

In our implementation we use a nearest-neighbor matchgate circuit generated by layers of single-mode rotations and two-mode couplings, such as
\[
e^{i\alpha Z_i},\qquad e^{i\beta X_iX_{i+1}},
\]
together with discrete Pauli reflections when needed. 

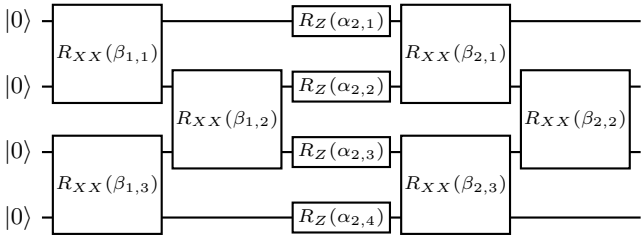
\begin{figure}[H]
    \centering
    \begin{tikzpicture}[thick,scale=.72]
        \foreach \y in {0, -1.2, -2.4, -3.6} {
            \draw (2, \y) -- (13, \y);
            \node[left] at (2, \y) {$|0\rangle$}; 
        }

        \node[draw, rectangle, fill=white, inner sep=2pt,font=\scriptsize] at (7.5, 0) {$R_Z{(\alpha_{2,1})}$};
        \node[draw, rectangle, fill=white, inner sep=2pt,font=\scriptsize] at (7.5, -1.2) {$R_Z{(\alpha_{2,2} )}$};
        \node[draw, rectangle, fill=white, inner sep=2pt,font=\scriptsize] at (7.5, -2.4) {$R_Z{(\alpha_{2,3})}$};
        \node[draw, rectangle, fill=white, inner sep=2pt,font=\scriptsize] at (7.5, -3.6) {$R_Z{(\alpha_{2,4})}$};

\draw[fill=white] (2.2, 0.3) rectangle (4.2, -1.5)
    node[pos=0.5, font=\scriptsize] {$R_{XX}(\beta_{1,1})$};

\draw[fill=white] (2.2, -2.1) rectangle (4.2, -3.9)
    node[pos=0.5, font=\scriptsize] {$R_{XX}(\beta_{1,3})$};

\draw[fill=white] (4.4, -0.9) rectangle (6.4, -2.7)
    node[pos=0.5, font=\scriptsize] {$R_{XX}(\beta_{1,2})$};

\draw[fill=white] (8.6, 0.3) rectangle (10.6, -1.5)
    node[pos=0.5, font=\scriptsize] {$R_{XX}(\beta_{2,1})$};

\draw[fill=white] (8.6, -2.1) rectangle (10.6, -3.9)
    node[pos=0.5, font=\scriptsize] {$R_{XX}(\beta_{2,3})$};

\draw[fill=white] (10.8, -0.9) rectangle (12.8, -2.7)
    node[pos=0.5, font=\scriptsize] {$R_{XX}(\beta_{2,2})$};
    \end{tikzpicture}
    \caption{An example of matchgate ansätze}
\end{figure}

This family of quantum circuit has a property of humming weight parity protection. A more specified ansätze with fixed humming weight is given by $XY$-mixer $X_{i}X_{j}+Y_{i}Y_{j}$, and a near-neighbor case when $j=i+1$ lies in the class of matchgate composed by $R_Z, R_{XX}$ because $e^{i\frac{\theta}{2}(X_iX_{i+1}+Y_iY_{i+1})}$ can be decomposed into $R_Z$ and $R_{XX}$.
They are more suitable for Boolean optimization problems with constraints $\sum_i x_i\equiv 0/1(\text{mod}\ 2)$ or $\sum_i x_i=K$.

Another ansätze with more flexible humming weight is obtained by a global Hadamard transformation, yielding gates of the form
\[
e^{i\alpha X_i},\qquad e^{i\beta Z_iZ_{i+1}}.
\]

More generally, any circuit satisfying the Majorana linear transformation condition belongs to the fermionic Gaussian class.

\emph{Covariance matrix simulation.}
The key computational advantage arises from the covariance matrix representation. For a fermionic Gaussian state define
\[
\Gamma_{pq}=\frac{i}{2}\langle[c_p,c_q]\rangle.
\]
If a matchgate induces the Majorana orthogonal transformation \(R\), then the covariance matrix updates as
\[
\Gamma\mapsto R\Gamma R^T.
\]
For a circuit with layers \(R_1,\ldots,R_L\), the total transformation is
\[
R_{\rm tot}=R_LR_{L-1}\cdots R_1,
\]
and
\[
\Gamma_{\rm out}=R_{\rm tot}\Gamma_{\rm in}R_{\rm tot}^T.
\]
The covariance matrix has dimension \(2n\times 2n\), avoiding storage of the full \(2^n\)-dimensional state vector.

Observable evaluation follows from Wick's theorem. For an even product of Majorana operators,
\[
\langle c_{p_1}c_{p_2}\cdots c_{p_{2k}}\rangle=\operatorname{Pf}(M),
\]
where \(M\) is the antisymmetric submatrix determined by \(\Gamma_{\rm out}\). For definite-parity Gaussian states, odd Majorana correlators vanish. Since \(Z_i=-ic_{2i-1}c_{2i}\), any \(Z\)-string from a diagonal Boolean cost term becomes an even Majorana product and can be evaluated by a Pfaffian.

Crucially, for a cost Hamiltonian
\[
H=\sum_{j=1}^m w_j H_j,
\]
we do not simulate the circuit separately for each \(H_j\). For fixed \(\theta\), the circuit is propagated once to obtain \(\Gamma_{\rm out}(\theta)\). All local terms \(\langle H_j\rangle\) are then read from the same covariance matrix and summed:
\[
L(\theta)=\sum_{j=1}^m w_j\langle H_j\rangle_\theta.
\]
This ``one state propagation, many observable readouts'' mechanism is essential for large Boolean instances with many local constraints.

\paragraph{Numerical results.} We benchmark on a series of problem called polynomial unconstrained binary operations (PUBO), especially widely-used QUBO with degree 2. The results of gradient descent reach more than 99.5\% of the best energy given from a classical method (20*500 sweeps of simulated annealing), while the circuit depth required is logarithmic.

\begin{table}[H]

\label{tab:convergence}
\small
\setlength{\tabcolsep}{1.5pt}
\begin{ruledtabular}
\begin{tabular}{c|cccccc}
\(n\)&200 & 1000 & 2000 & 3000 & 4000 & 5000 \\
\hline
Terms&200 & 1000 & 2000 & 3000 & 4000 & 5000 \\
Depth&93& 120 & 132 & 138 & 144 & 147 \\
Grads & 81.63&  116.95 &  135.44 & 146.88  & 156.48 & 164.46 \\
Score & 99.647\% & 99.673\%  & 99.702\%  & 99.684\%  & 99.670\% & 99.682\% \\
\end{tabular}
\end{ruledtabular}
\caption{Convergence quality and cost for QUBOs.}
\end{table}

\begin{figure}
    \centering
    \includegraphics[width=\linewidth]{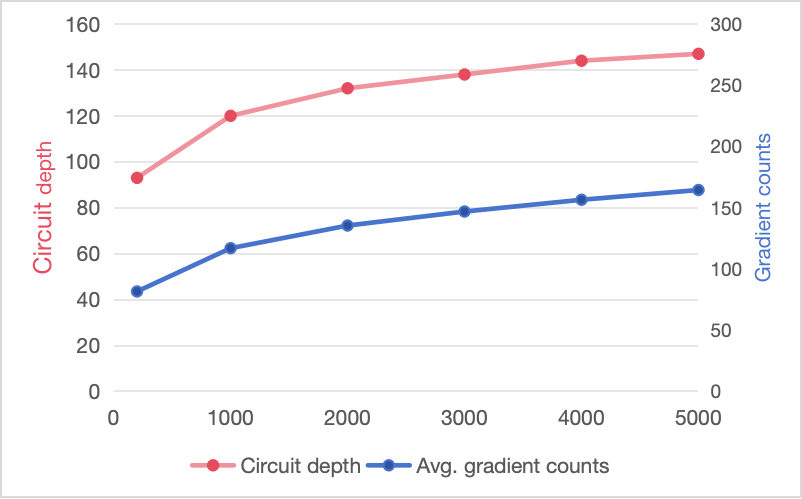}
    \caption{Scaling law: QUBO}
    \label{fig:qubo-scaling}
\end{figure}

To further explore the potential for quantum advantage, we extend our investigation to denser and more complex problems. We test on max constraint satisfaction problems, especially ones with planted solutions, so that optimal values are known to be number of constraints. The most classical example is SAT, the first problem proved to be NP-complete \cite{cook1971complexity}, and therefore has the longest history and fruitful legacies to solve. Each clause is encoded as a local diagonal cost term, with satisfied clauses assigned low energy and violated clauses assigned high energy. We consider random 3-SAT instances with clause-to-variable ratio near-threshold, around \(4.267\), where is empirically hardest to solve.

The optimized energies remain extremely close to the optimum. Across all tested sizes, the average approximation ratio exceeds 99.5\% and remains nearly the same. The quantum circuit is set to be $\text{round}(4\log_2n)$ repeated double layers of neighboring $R_{XX}$ and a layer of $R_{Z}$.

\begin{table}[H]
\small
\setlength{\tabcolsep}{1.5pt}
\begin{ruledtabular}
\begin{tabular}{c|cccccc}
\(n\) & 200 & 1000 & 2000 & 3000 & 4000 & 5000 \\
\hline
Clauses & 853 & 4267 & 8534 & 12801 & 17068 & 21335 \\
Depth & 93 & 120 & 132 & 138 & 144 & 147 \\
Grads & 136.64 & 142.18 & 143.60 & 145.88 & 146.88 & 146.20 \\
Energy & 849.84 & 4250.44 & 8500.10 & 12752.68 & 17002.89 & 21253.33 \\
Ratio & 99.630\% & 99.612\% & 99.603\% & 99.623\% & 99.619\% & 99.617\%
\end{tabular}
\end{ruledtabular}
\caption{Convergence quality and cost for 3-SATs.}
\label{tab:convergence_3sat}
\end{table}

Note that Adam causes oscillations at the end, so the gradient counts are fluctuated and therefore not monotonous.

After a few steps of post-processing with methods like WalkSAT, the outcome sample strings of the optimized variational quantum circuit can be fixed to true solution of the 3-SAT problem, obtaining global optimum. For small instances, an optimal solution is reached within a few trials of gradients without post-processing. 

For 4-SATs and 5-SATs, the logarithmic scaling law appears the same:
\begin{table}[H]
\small
\setlength{\tabcolsep}{1.5pt}
\begin{ruledtabular}
\begin{tabular}{c|cccccc}
\(n\) & 200 & 1000 & 2000 & 3000 & 4000 & 5000 \\
\hline
Clauses & 1986 & 9931 & 19862 & 29793 & 39724 & 49655 \\
Depth & 93 & 120 & 132 & 138 & 144 & 147 \\
Grads & 132.72 & 137.20 & 139.46 & 140.54 & 140.80 & 141.76 \\
Energy & 1977.11 & 9886.37 & 19770.55 & 29656.70 & 39543.09 & 49427.97 \\
Ratio & 99.552\% & 99.551\% & 99.540\% & 99.542\% & 99.545\% & 99.543\%
\end{tabular}
\end{ruledtabular}
\caption{Convergence quality and cost for 4-SATs.}
\label{tab:convergence_4sat}
\end{table}

\begin{table}[H]
\small
\setlength{\tabcolsep}{1.5pt}
\begin{ruledtabular}
\begin{tabular}{c|cccccc}
\(n\) & 200 & 1000 & 2000 & 3000 & 4000 & 5000 \\
\hline
Clauses & 4223 & 21117 & 42234 & 63351 & 84468 & 105585 \\
Depth & 93 & 120 & 132 & 138 & 144 & 147 \\
Grads & 127.70 & 132.50 & 134.04 & 135.08 & 136.64 & 137.14 \\
Energy & 4208.80 & 21044.63 & 42091.54 & 63140.51 & 84183.52 & 105233.47 \\
Ratio & 99.664\% & 99.657\% & 99.663\% & 99.668\% & 99.663\% & 99.667\%
\end{tabular}
\end{ruledtabular}
\caption{Convergence quality and cost for 5-SATs.}
\label{tab:convergence_5sat}
\end{table} 

Our matchgate method gives a quantum-inspired algorithm which outperforms some classical algorithms. Classical MaxSAT methods include SAT-based linear search (LSU),
core-guided optimization (RC2), implicit hitting sets (MaxHS),
and stochastic local search~\cite{bacchus2021maxsat,
martins2014openwbo,ignatiev2019rc2,davies2011maxhs}.
Strong local-search baselines include probSAT and YalSAT for
random satisfiable SAT, and SATLike and NuWLS for (weighted)
partial MaxSAT~\cite{biere2017yalsat,lei2018satlike,chu2023nuwls}.
Recent approaches combine local search, SAT-based reasoning,
and mathematical programming, including SLS-enhanced core-boosted
search, UWrMaxSAT~2.0, and Aperture~\cite{lubke2025sls,
piotrow2026uwrmaxsat,nadel2026aperture}.
Our matchgate method outperforms the tested LSU and
SATNet-based semidefinite relaxation configurations on all ten
paired planted 5-SAT instances at each of $n=200,1000,5000$,
even when these baselines receive twice the corresponding MG
training time (Table~\ref{tab:mg-classical-5sat}).
At $n=5000$, MG achieves 345.40 mean unsatisfied clauses,
compared with 1677.70 for LSU and 1167.60 for SATNet-SDP
at the doubled budget, without classical postprocessing.
These advantages are specific to the tested configurations;
our separate experiments do not establish an advantage over
YalSAT. Hardware and timing differences also preclude an
end-to-end speedup claim.

\begin{table}[H]
\small
\setlength{\tabcolsep}{4pt}
\begin{ruledtabular}
\begin{tabular}{l|ccc}
$n$ & 200 & 1000 & 5000 \\
\hline
Instances & 10 & 10 & 10 \\
Clauses & 4223 & 21117 & 105585 \\
Mean training budget (s) & 7.24 & 36.81 & 1220.30 \\
MG & 15.50 & 71.30 & 345.40 \\
\hline
LSU ($1\times$) & 58.60 & 362.90 & 1677.70 \\
LSU ($2\times$) & 54.00 & 362.90 & 1677.70 \\
SATNet-SDP ($1\times$) & 27.50 & 206.10 & 1169.30 \\
SATNet-SDP ($2\times$) & 26.50 & 204.10 & 1167.60 \\
\hline
\end{tabular}
\end{ruledtabular}
\caption{Mean unsatisfied clauses on the first ten original planted
5-SAT instances per size (lower is better). Classical budgets are one
or two times each instance's measured MG training time; MG sampling
is additional. MG uses Adam with the bias-correction counter fixed
at one. We report the best raw
sample without postprocessing: 30 samples at $n=200,1000$ and four at
$n=5000$. LSU results are reused. SATNet-SDP uses the official CUDA
coordinate-descent kernel with our fixed-CNF adapter and rounding
policy, not a trained network; its search clock excludes setup and
validation. LSU includes formula parsing. GPU/CPU and
timing differences preclude an end-to-end speedup claim.
This table reports selected LSU and SATNet-SDP comparisons,
not an exhaustive comparison with classical solvers.}
\label{tab:mg-classical-5sat}
\end{table}

\emph{Discussions.}
Our numerical results implement the scaling law of variational quantum algorithm is that, logarithmic circuit depth is sufficient to remain high convergence rate of combinatorial optimizations. Within classical simulation regime, we test matchgate which is the variational quantum circuits with lowest computational complexity, while in general universal quantum circuits are linear combinations of matchgates \cite{feng2025quon}. It is also useful to achieve on various platforms including superconductors ion traps and Rydberg atoms, since matchgate is free evolution of 1+1D transversal Ising models. Parallel classical simulations are useful for calibrations of quantum hardwares. We also expect more general ansätze to be more expressive, admitting better optimization results.

The matchgate ansätze also gives a classical heuristic algorithms which cooperates well with existing algorithms such as classical annealings. It's interesting to study the property of this quantum-inspired heuristic algorithm in similarities and extensive strength compared with other classical heuristic algorithms and optimization methods.

\paragraph{Acknowledgements.} The authors thank Yuguo Shao for valuable discussions. Patent applications related to this work have been filed. Commercial use of the technology described herein may require a license from the patent owner. Z.L. acknowledges support from the Beijing Natural Science Foundation (Grant No. Z220002).
\bibliographystyle{unsrt}
\bibliography{sample}

\end{document}